\documentclass[twocolumn,showpacs,preprintnumbers,a4,prl]{revtex4}
\usepackage{graphicx}
\usepackage{dcolumn}
\usepackage{amsmath}
\usepackage{amssymb}
\usepackage{bm}

\begin{document}

\title{Relaxation measurements in the regime of the second magnetization peak in Nb films}
\author{D. Stamopoulos \footnote[1]{Corresponding author. Tel.: +30-2106503321, fax: +30-2106519430, e-mail address:
densta@ims.demokritos.gr (D. Stamopoulos)} and D. Niarchos}

\affiliation{Institute of Materials Science, NCSR "Demokritos",
153-10, Aghia Paraskevi, Athens, Greece.}

\date{\today}

\begin{abstract}
We report on magnetic measurements as a function of field,
temperature and time (relaxation) in superconducting Nb films of
critical temperature $T_c=9.25$ K. The magnetic measurements as a
function of field exhibited a ''second magnetization peak''
(''SMP'') which in general is accompanied by thermomagnetic
instabilities (TMIs). The lines $H_{\rm ''smp''}(T)$ where the
''SMP'' occurs and the $H_{\rm fj}(T)$ where the first flux jump
in the virgin magnetization curves is observed, end at a
characteristic point $(T_o,H_o)\approx(7.2$ K,$80$ Oe). Relaxation
measurements showed that for $T<T_o\approx7.2$ K the activation
energy $U_o$ and the normalized relaxation rate $S$ exhibit
non-monotonic behavior as a function either of temperature or
field. The extrema observed in $U_o$ and $S$ coincide with the
points $H_{\rm on}(T)$ or $H_{\rm ''smp''}(T)$ of the ''SMP''. In
the regime $T>T_o\approx7.2$ K both $U_o$ and $S$ present a
conventional monotonic behavior. These results indicate that the
''SMP'' behavior observed in our Nb films is promoted by the
anomalous relaxation of the magnetization.

\end{abstract}


\maketitle

\section{Introduction}

Recent magnetic studies in Nb films
\cite{Esquinazi99,Kopelevich98,Stamopoulosnew} reported the
existence of a structure that reminisces of the second
magnetization peak (SMP) which is usually observed in high-$T_c$
superconductors
\cite{Zhukov95,Stamopoulos01,Stamopoulos02,Stamopoulos03,Sun00,Kupfer94,Janossy96,Pissas99,Pissas00,Sun02,Chowdhury03,Abulafia96,Miu00,KhaykovichPRL,Kokkaliaris00,Giller97}.
These studies concluded that for the case of Nb films the ''SMP''
is probably motivated by thermomagnetic instabilities (TMIs) that
occur in the low-temperature regime, far below the upper-critical
field \cite{Esquinazi99,Kopelevich98,Stamopoulosnew}. From the
theoretical point of view, the understanding of the fundamental
differences between the ''SMP'' observed in Nb films and in
high-$T_c$ superconductors is of great importance. In addition,
the detailed experimental study of the TMIs that accompany the
''SMP'' in Nb films is also very important, because these
undesirable flux jumps constitute a serious limitation for basic
practical applications.

Today, a detailed study of the dynamic response of vortices in the
regime of the ''SMP'' in Nb films is lacking. In this work we
present detailed magnetic data and especially we study the
relaxation of vortices in sputtered thick Nb films that exhibit
the ''SMP''. We systematically recorded the lines $H_{\rm
''smp''}(T)$ where the ''SMP'' occurs, $H_{\rm fj}(T)$ where the
first flux jump takes place in the virgin magnetization curves and
$H_{\rm fp}(T)$ where the first peak is observed. These three
lines connect at a characteristic point $(T_o,H_o)\approx(7.2$
K,$80$ Oe). For $T>T_o$ no flux jumps are observed and the
magnetic curves are absolutely smooth. Interestingly, below a
second characteristic point $(T_1,H_1)\approx(6.2$ K,$1600$ Oe)
the line $H_{\rm ''smp''}(T)$ changes slope. In the regime
$T_1<T<T_o$ TMIs are pronounced and the ''SMP'' faints as the
characteristic temperature $T_o$ is approached. In contrast, for
$T<T_1$ TMIs are strongly suppressed. As a consequence, in this
regime the magnetic curves are almost regular and the ''SMP''
strongly resembles the respective feature observed in high-$T_c$
compounds. Detailed relaxation measurements, showed that the
activation energy $U_o$ (normalized relaxation rate $S$) exhibits
a maximum (minimum) as a function of temperature or field for
$T<T_o\approx7.2$ K. The observed extrema are located exactly at
the $H_{\rm ''smp''}(T)$ or $H_{\rm on}(T)$ points. In contrast,
in the regime $T>T_o\approx7.2$ K a monotonic behavior is observed
for both $U_o$ and $S$. Although our results indicate that in Nb
films the ''SMP'' feature is simply motivated by an anomalous
relaxation mechanism, the similarities of the observed behavior
with the behavior observed in high-$T_c$ compounds calls for
further theoretical and experimental work.

\section{Preparation of the films and Experimental details}

The samples of Nb were sputtered on Si $[001]$ substrates as they
were annealed during the deposition. Details on the preparation of
the films are presented elsewhere \cite{Stamopoulosnew}. The
critical temperature of the film under discussion is $T_c=9.25$ K,
equal to the one of high purity single crystals
\cite{Finnemore66}. The residual resistance ratio is R($300$
K)/R($10$ K)=$6.8$ ( $\varrho _n \approx 10$ $\mu \Omega$cm just
above $T_c$). The thickness of the film is $7700$ {\AA}. Our
combined x-ray diffraction (XRD) and transmission electron
microscopy (TEM) data revealed that by annealing the films during
deposition, a larger mean size of the grains is produced, which in
this case is $930$ {\AA}. Furthermore, the grains of the annealed
films are oriented, in some degree, with $[110]$ direction
perpendicular to the film's surface. Our magnetic measurements
either as a function of field, temperature or time were performed
by means of a commercial SQUID magnetometer (Quantum Design). In
all magnetic data presented below, the magnetic field was always
normal to the surface of the film (${\bf H}\parallel{\bf c}$). All
the magnetic measurements were performed under zero field cooling
(ZFC), so that we always obtained the virgin magnetization curves.
In relaxation measurements the magnetization of the sample was
recorded for a time window of $100$ sec$<t<4200$ sec. We examined
the whole temperature-magnetic-field regime accessible by our
SQUID ($H_{\rm dc}<55$ kOe, $T>1.8$ K).

\section{Experimental results and discussion}

\subsection{TMIs and the ''SMP'' in magnetic measurements}

\begin{figure}[tbp] \centering%
\includegraphics[angle=0,width=8cm]{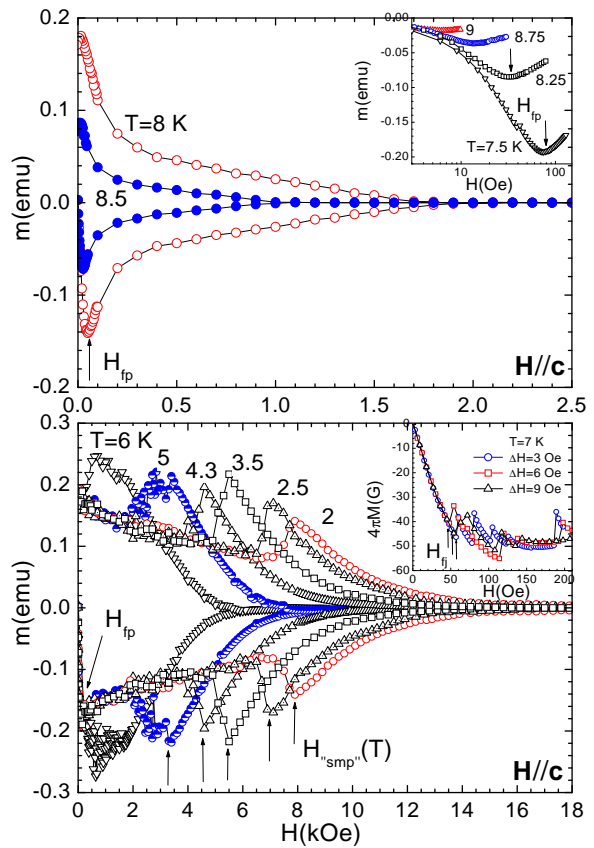}%
\caption {Isothermal magnetic m(H) measurements for $T=8$ and
$8.5$ K, above $T_o\approx7.2$ K (upper panel), and for $T=6, 5,
4.3, 3.5, 2.5$ and $2$ K, below $T_o\approx7.2$ K (lower panel).
For $T>T_o\approx7.2$ K the magnetic curves are smooth, while for
$T<T_o\approx7.2$ K they are accompanied by TMIs, presenting a
''SMP'' feature at points $H_{\rm ''smp''}$. At very low
temperatures the TMIs are strongly suppressed and the ''SMP'' is
evident. In the inset of the upper panel we present magnetic
measurements in low magnetic fields, at various temperatures, for
the determination of the first peak field $H_{\rm fp}$. In the
inset of the lower panel we focus on the flux jumps that occurred
at $T=7$ K, for the determination of the first flux jump field
$H_{\rm fj}$. Magnetic curves for various sweep rates of the
applied field are presented. In all cases the magnetic field was
normal to the surface of the film ${\bf H}\parallel{\bf c}$.}
\label{b1}%
\end{figure}

We start the presentation of our experimental results with the
magnetic measurements as a function of field. Figures \ref{b1}(a)
and \ref{b1}(b) present data in the high (upper panel) and in the
low-temperature regime (lower panel). We see that close to $T_c$
(upper panel) the magnetic curves are smooth, while the loops at
low temperatures (lower panel) are anomalous. More specifically,
below the characteristic temperature $T_o=7.2$ K a ''noisy'' first
peak shows up which is accompanied by strong TMIs. As the
temperature is gradually reduced this broad peak transforms into
two separate peaks: a first maximum which occurs in low-field
values, and a distinct second peak which is placed in high
magnetic fields. We note that the magnetic loops performed in the
interval $T_1\approx 6.2$ K$<T<T_o\approx 7.2$ K exhibited strong
TMIs. In this regime the ''SMP'' is hardly observed. In contrast,
for even lower temperatures $T<T_1\approx 6.2$ K, the ''SMP'' is
pronounced and the TMIs are strongly suppressed. As a result the
magnetic curves are more regular. The overall behavior resembles
the SMP observed usually in high-$T_c$ superconductors
\cite{Zhukov95,Stamopoulos01,Stamopoulos02,Stamopoulos03,Sun00,Kupfer94,Janossy96,Pissas99,Pissas00,Sun02,Chowdhury03,Abulafia96,Miu00,KhaykovichPRL,Kokkaliaris00,Giller97}.
For this reason we ''loosely'' refer to these characteristic
fields as $H_{\rm ''smp''}$. However, there are noticeable
differences between the magnetic behaviors  observed in our Nb
film and in high-$T_c$ superconductors. In our case, below the
''SMP'' ($H<H_{\rm ''smp''}$) the response is ''noisy'', while
above it ($H>H_{\rm ''smp''}$) we observed smooth magnetic curves.
In contrast, in {\it point disordered} high-$T_c$ superconductors
the magnetization curves are smooth, both below and above the SMP.
In addition, the peak value of the $H_{\rm ''smp''}$ slightly
decreases as we lower the temperature. In high-$T_c$
superconductors for lower temperatures the peak value of the SMP
strongly increases. Finally, in our Nb film the resulting line
$H_{\rm ''smp''}(T)$ ends on the first peak line $H_{\rm fp}(T)$
at $T_o/T_c\approx 0.78$ (see Fig. \ref{b3} below), while in
high-$T_c$ superconductors the respective line $H_{\rm smp}(T)$
ends near the irreversibility/melting line, or extends almost up
to the critical temperature. In the respective inset of the upper
(lower) panel we present measurements at low magnetic fields, for
the demonstration of the first peak field $H_{\rm fp}$ (first flux
jump field $H_{\rm fj}$).

\begin{figure}[tbp] \centering%
\includegraphics[angle=0,width=8cm]{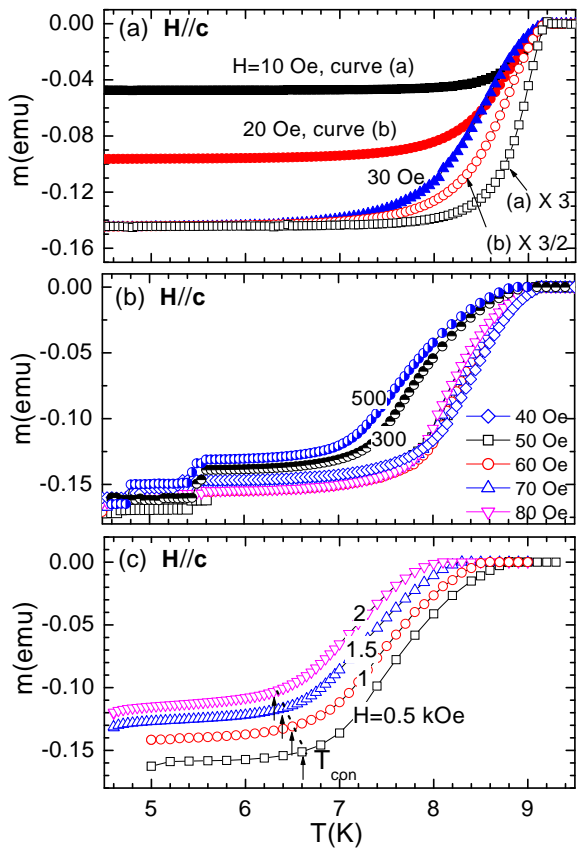}
\caption {Isofield magnetic measurements m(T) under various
magnetic fields (a) $H=10,20$ and $30$ Oe, (b) $H=40-500$ Oe and
(c) $H=0.5-2$ kOe. In very low fields (Fig. (a)) the magnetic
moment exhibits the usual behavior observed in the Meissner state.
In intermediate fields (Fig. (b)) we enter the regime where TMIs
are observed (see also Fig. \ref{b3} below). At even higher fields
(Fig. (c)) we observe that the magnetization is almost constant
below the characteristic temperatures $T_{\rm con}$, while for
$T>T_{\rm con}$ the m(T) curves start to decrease rapidly. The
points $(T_{\rm con},H)$ coincide with the points $(T,H_{\rm
ep}^{\rm TMIs})$ where the last TMI is observed in our m(H) curves
(see Fig. \ref{b3} below). In all cases the magnetic field was
normal to the surface of the film ${\bf H}\parallel{\bf c}$.}
\label{b2}%
\end{figure}%

In Figs. \ref{b2}(a)-(c) we present magnetic data as a function of
temperature in various magnetic fields. In Fig. \ref{b2}(a) we
present data at low applied fields of the order of the first
critical field. In addition to the raw data we also present two
more curves that resulted after multiplying the original curves
(a) and (b) (referring to $H=10$ and $20$ Oe) by a factor of $3$
and $3/2$ respectively. We clearly see that the behavior of the
low temperature regime is the one expected in the Meissner state
since the resulting curves coincide with the curve that refers to
field $H=30$ Oe. In Fig. \ref{b2}(b) we present data for
intermediate field values. These data exhibit TMIs in the low
temperature regime. Finally, in Fig. \ref{b2}(c) we show
representative results for comparatively high magnetic fields. We
see that until some characteristic temperature $T_{\rm con}$ the
magnetic moment is almost constant, while above it ($T>T_{\rm
con}$) is strongly reduced. The characteristic points $(T_{\rm
con},H)$ that come from these isofield measurements as a function
of the temperature, coincide with the boundary above which TMIs
are no longer observed $(T,H_{\rm e p}^{\rm TMIs})$ as determined
from isothermal magnetic loop measurements (see Fig. \ref{b3}).

\begin{figure}[tbp] \centering%
\includegraphics[angle=0,width=9cm]{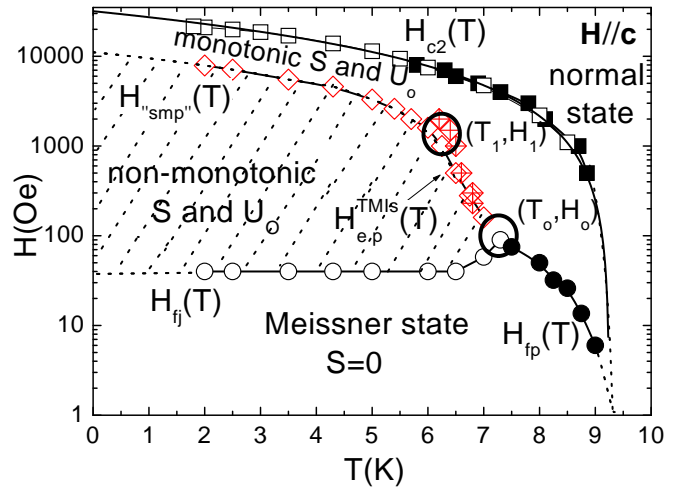}
\caption { Low-field part of the resulting phase diagram for the
Nb film (${\bf H}\parallel{\bf c}$). Presented are the lines
$H_{\rm fp}(T)$ (solid circles), the $H_{\rm fj}(T)$ (open
circles), the ''SMP'' $H_{\rm ''smp''}(T)$ (rhombi) and the
upper-critical field $H_{\rm c2}(T)$ (squares). The rhombi with
crosses come from isofield magnetic measurements as a function of
temperatures (see Fig. \ref{b2}). We notice that the part of the
''SMP'' line that lies in the regime $T>T_1\approx 6.2$ K refers
to the boundary $H_{\rm ep}^{\rm TMIs}(T)$ where the last TMI is
observed in magnetic curves. In the regime of TMIs (shaded area) a
non-monotonic behavior is observed for $S$ and $U_o$, while
outside this regime both $S$ and $U_o$ exhibit a regular monotonic
behavior. Below the $H_{\rm fj}(T)$ line the magnetization doesn't
relax.}
\label{b3}%
\end{figure}%

The resulted ''phase diagram'' of vortex matter for the Nb film is
presented in Fig. \ref{b3}. Presented are the lines $H_{\rm
''smp''}(T)$ (rhombi), the first peak field $H_{\rm fp}(T)$ (solid
circles) and the first flux jump field $H_{\rm fj}(T)$ (open
circles). In addition, the upper-critical fields $H_{\rm c2}(T)$
(squares) are presented. The upper-critical points were determined
by comparative magnetic and magnetoresistance measurements (not
shown here) \cite{Stamopoulosnew}. As we see the $H_{\rm fp}(T)$,
$H_{\rm fj}(T)$ and $H_{\rm ''smp''}(T)$ lines connect at a
characteristic point $(T_o,H_o)\approx (7.2\ {\rm K}, 80\ {\rm
Oe})$. We clearly see that in the temperature range $T>T_o=7.2$ K,
where the line $H_{\rm fj}(T)$ tends to overcome the first peak
line $H_{\rm fp}(T)$, TMIs are no longer observed. In addition, in
the regime $T<T_o=7.2$ K the experimental data for $H_{\rm fj}(T)$
can't be described by any of the proposed theoretical curves
\cite{Stamopoulosnew}. Interestingly, at $(T_1,H_1)\approx (6.2\
{\rm K}, 1600\ {\rm Oe})$ the line $H_{\rm ''smp''}(T)$ changes
slope. We must underline that as we move close to the point
$(T_o,H_o)\approx (7.2\ {\rm K}, 80\ {\rm Oe})$ the ''SMP'' is
hardly observed. As a result, the part of the $H_{\rm ''smp''}(T)$
line that lies above the point $T_1\approx 6.2$ K, simply refer to
the boundary $H_{\rm ep}^{\rm TMIs}(T)$ above which TMIs are not
observed. Below $T_1\approx 6.2$ K the $H_{\rm ''smp''}(T)$ line
is placed in high fields, while the $H_{\rm fj}(T)$ line takes the
constant value $H_{\rm fj}(T<T_1)=40$ Oe. This experimental
result, of a temperature independent $H_{\rm fj}(T)$ line, is in
contrast to theoretical predictions
\cite{Swartz68,Wipf6791,Mints81,Mints96} that treat the $H_{\rm
fj}(T)$ line as a simple boundary above which TMIs occur
\cite{Stamopoulosnew}.

\subsection{Relaxation measurements in the regime of the ''SMP''}

\begin{figure}[tbp] \centering%
\includegraphics[angle=0,width=9cm]{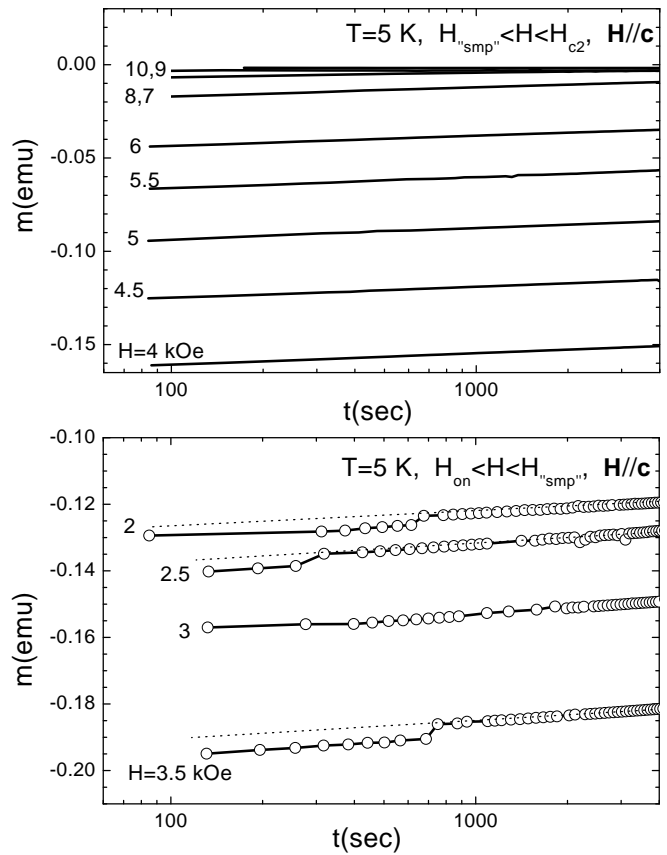}
\caption {Magnetic measurements m(t) as a function of time under
various magnetic fields $H=2-10$ kOe, at $T=5$ K. In the upper
panel we present data in the field regime $H_{\rm
''smp''}<H<H_{\rm c2}$, while in the lower panel we present
respective data for $H_{\rm on}<H<H_{\rm ''smp''}$. Clearly, the
magnetic moment exhibits a logarithmic relaxation. In the lower
panel we see that a few jumps are present, as these measurements
were performed in the regime of TMIs (see Fig. \ref{b3} above). In
all cases the magnetic field was normal to the surface of the film
${\bf H}\parallel{\bf c}$.}
\label{b4}%
\end{figure}

\begin{figure}[tbp] \centering%
\includegraphics[angle=0,width=8cm]{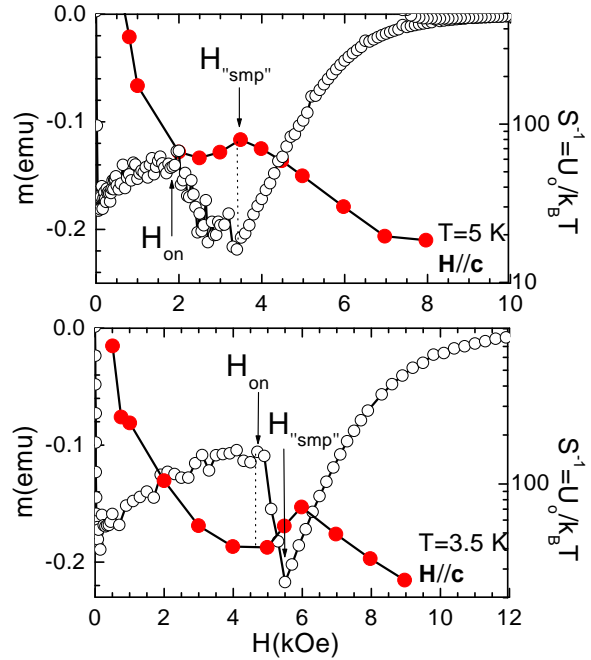}
\caption {The ascending branch of the magnetic measurement as a
function of field and the estimated activation energy $U_o/k_BT$
in various magnetic fields, at $T=5$ K (upper panel) and  at
$T=3.5$ K (lower panel). We see that $U_o/k_BT$ presents a
non-monotonic behavior. In intermediate temperatures ($T=5$ K) the
local maximum of $U_o/k_BT$ is located at exactly the peak point
$H_{\rm ''smp''}$ of the ''SMP''. At lower temperatures ($T=3.5$
K) it is the local minimum of $U_o/k_BT$ that coincides with the
onset point $H_{\rm on}$ of the ''SMP'', while the local maximum
of $U_o/k_BT$ is placed above the $H_{\rm ''smp''}$ point. In all
cases the magnetic field was normal to the surface of the film
${\bf H}\parallel{\bf c}$.}
\label{b6}%
\end{figure}

By measuring the magnetic moment as a function of time (relaxation
measurements) new information may be gained about the pinning
mechanism. Furthermore, additional information at (very short or
long) times not accessible by a realistic experiment may be
obtained, by extrapolating the measured curves. Of course, this is
permitted under the assumption that the relaxation mechanism
doesn't change. Our main aim was the study of the dynamic behavior
of vortices in the regime of the ''SMP''. To this end, we
performed detailed relaxation measurements at temperatures
$T<T_o$. In addition, relaxation measurements were performed at
temperatures $T>T_o$ where the ''SMP'' is not observed, in order
to be compared with the respective ones at $T<T_o$. In Fig.
\ref{b4} we show representative data at $T=5$ K for various fields
$H=2-10$ kOe. In the upper (lower) panel we present results in the
field regime above (below) the ''SMP'' (see Fig. \ref{b6}(a)
below). It is clear that the magnetic moment relaxes according to
a logarithmic law, as this is evident by its linear variation on a
semilogarithmic plot. The curves of the upper panel are smooth and
their slope changes slowly as the upper-critical field is
approached. In the lower panel, where the measurements were
performed in the regime of TMIs (see Fig. \ref{b6}(a) below),
small flux jumps are observed. These jumps are restricted in short
measuring times. Interestingly, the slope of the curves below and
above each jump is almost the same. Nevertheless, in such cases,
in the processing of the data we took into account only the part
of the curves placed above the jump.

Since we study a low-$T_c$ superconductor which exhibits
logarithmic relaxation we used the classical Anderson-Kim model
\cite{Anderson62} for the estimation of the activation energy
$U_o$ and the respective normalized relaxation rate $S$. For
Anderson-Kim flux creep the potential barrier that vortices
experience is linear on the applied current according to the
relation $U(j)=U_o(1-j/j_c)$. It comes out that the bulk screening
currents that exist in the superconductor due to the pinning of
vortices, relax according to a logarithmic law. The relaxation of
the magnetization of a thin disk under a perpendicular magnetic
field has been also treated analytically in Ref.
\onlinecite{Gurevich94}. In that work it was pointed out that in
the regime of steady state relaxation (at times much higher than a
characteristic transient time) the relaxing magnetization is
independent of the initial conditions of the sweep rate of the
applied field. The result obtained for the steady state relaxation
is
\begin{equation}
m(t)=m_c-m_1\ln(\frac{t}{t_o}), \label{eq8}
\end{equation}
where $t_o\approx j_cdaS/c^2E_c$, $m_c=\pi a^3j_cd/3c$ and
$m_1=Sm_c$. In the above quantities $t_o$ is a characteristic
microscopic attempt time, $d$ and $a$ are the thickness and the
''diameter'' of the sample respectively and finally, $E_c$ is a
threshold value of the electric field. By fitting the relaxation
data we estimated the related pinning parameters $U_o$ and $S$. In
our fitting procedures we assumed that the attempt time is
$t_o\in[10^{-3}-10^{-6}]$ sec.

\begin{figure}[tbp] \centering%
\includegraphics[angle=0,width=8cm]{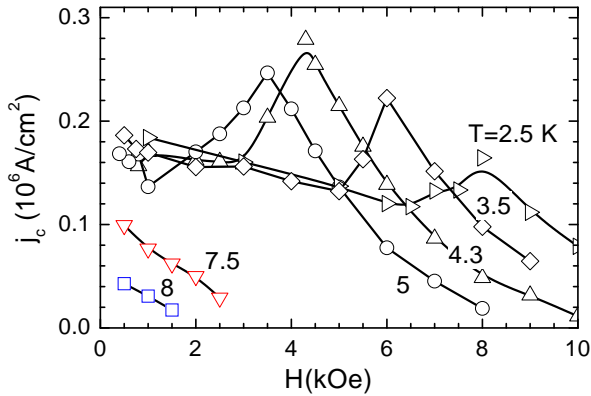}
\caption {Estimated critical current density at various
temperatures $T=2.5, 3.5, 4.3$ and $5$ K (below $T_o\approx 7.2$
K) and at $T=7.5$ and $8$ K (above $T_o\approx 7.2$ K). For
$T>T_o\approx 7.2$ K $j_c$ is monotonic in the whole field regime,
while for $T<T_o\approx 7.2$ K is non-monotonic presenting extrema
as a function of field. In all cases the magnetic field was normal
to the surface of the film ${\bf H}\parallel{\bf c}$.}
\label{b5}%
\end{figure}%

Figure \ref{b5} presents the estimated critical current density
$j_c(H)$ at various temperatures. We clearly see that for
$T<T_o\approx 7.2$ K $j_c(H)$ presents a non-monotonic behavior,
while at $T>T_o\approx 7.2$ K a regular monotonic behavior is
recovered. From these data we see that the ''SMP'' feature is
present even for the ''unrelaxed'' magnetization (at very small
times of the order of $t_o\in[10^{-3}-10^{-6}]$ sec).

In Fig. \ref{b6} we present comparatively the magnetic loops and
the estimated activation energy $U_o$ for various magnetic fields
at $T=5$ K (upper panel) and $T=3.5$ K (lower panel). As we
clearly see at very low magnetic fields, $U_o$ takes very high
values (consequently $S$ takes very low values). When increasing
the field, $U_o$ exhibits a non-monotonic behavior. More
specifically, for $T=5$ K {\it $U_o$ presents a maximum at exactly
the $H_{\rm ''smp''}$ point}. At lower temperatures the maximum in
$U_o$ was shifted in higher fields. For example, at $T=3.5$ K the
maximum in $U_o$ is placed above the $H_{\rm ''smp''}$ point,
while the local minimum of {\it $U_o$ is located at the onset
point $H_{\rm on}$ of the ''SMP''.} The appearance of these
extrema in $U_o$ (and also in $S$) means that there is an
enhancement of the effective pinning in the regime of the ''SMP''.
In a recent article H. Kupfer et al. \cite{Kupfer94} discussed
relaxation as a possible origin of the SMP in
YBa$_2$Cu$_3$O$_{7+\delta}$ single crystals. They proposed that
the SMP could originate from a crossover between different
relaxation mechanisms. Faster relaxation in the low field regime
could lead to a suppression of the measured curves (that relax
during the measurement), while slower relaxation at higher fields
would restore higher values of the measured magnetization. As a
consequence a SMP could be obtained. Under this point of view, if
$j_c(H)$ was determined mainly by relaxation, a maximum of
$j_c(H)$ should correspond to a minimum in $S$ and a maximum in
$U_o$. In the experiments of Ref. \onlinecite{Kupfer94} the
minimum in $S$ (maximum in $U_o$) was observed at a characteristic
field $H^*(T)$, placed well below the maximum point $H_{\rm
smp}(T)$ of the SMP. As a result the relaxation origin of the SMP
in YBa$_2$Cu$_3$O$_{7+\delta}$ single crystals remained a doubtful
possibility. In our case, we see that both extrema of $U_o$ are
closely related to the onset $H_{\rm on}(T)$ and the maximum
$H_{\rm ''smp''}(T)$ of the ''SMP''. This could indicate that in
contrast to YBa$_2$Cu$_3$O$_{7+\delta}$, in our case the ''SMP''
feature could originate from an anomalous relaxation process as
the magnetic field increases. Soon after, Y. Abulafia et
al.\cite{Abulafia96} proposed that the anomalous behavior of $S$
and $U_o$ could be ascribed to a change in the dynamic response of
vortices at $H^*(T)<H_{\rm smp}(T)$. They proposed that for
$H<H^*(T)$ the response of vortex system was elastic, while for
$H>H^*(T)$ the increase of $S$ (decrease of $U_o$) should be
caused by plastic deformations of the vortex solid
\cite{Abulafia96}. Since then, many experimental works confirmed
this notion in YBa$_2$Cu$_3$O$_{7+\delta}$ and in other high-$T_c$
compounds as HgBa$_2$CuO$_{4+\delta}$ and
Tl$_2$Ba$_2$CaCu$_2$O$_8$
\cite{Kupfer94,Janossy96,Pissas99,Pissas00,Sun02,Chowdhury03}.
More specifically, in high-$T_c$ compounds for fields $H>H^*(T)$
the activation energy $U_o$ follows a negative power-law
$U_o\propto H^\nu$ with $\nu\in[-0.5,-0.9]$
\cite{Abulafia96,Pissas99,Miu00,Sun02,Chowdhury03}. In our case we
observed that above the ''SMP'' the activation energy may be
described by a power-law $U_o\propto H^\nu$ with $\nu\approx -2$.
This value is very high and probably inconsistent with the notion
of plastic vortex creep. The differences between the relaxation
process observed in Nb films and in high-$T_c$ compounds as
discussed above, may give additional evidence that the feature of
the second peak effect is motivated by different physical
mechanisms.

\begin{figure}[tbp] \centering%
\includegraphics[angle=0,width=8cm]{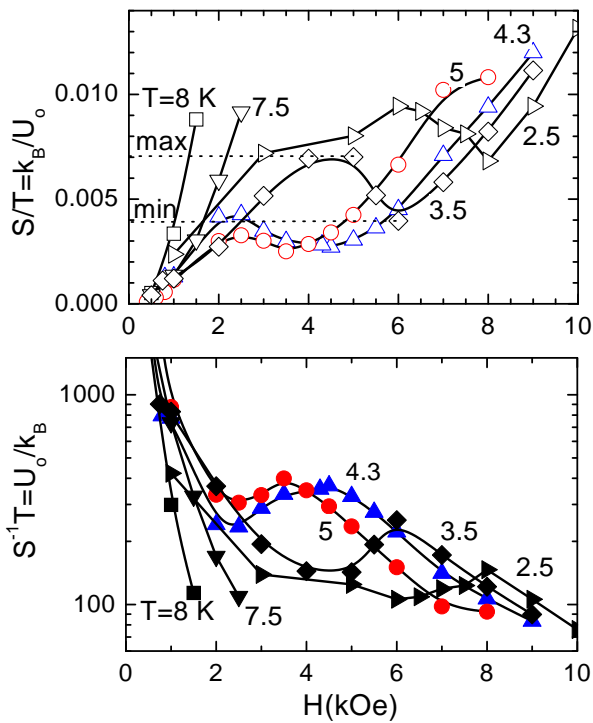}
\caption {Normalized relaxation rates $S/T$ (upper panel) and
activation energies $U_o/k_B$ (lower panel) at temperatures
$T=2.5, 3.5, 4.3$ and $5$ K (below $T_o\approx 7.2$ K) and at
$T=7.5$ and $8$ K (above $T_o\approx 7.2$ K). For $T>T_o\approx
7.2$ K the $S/T$ and $U_o/k_B$ are monotonic in the whole field
regime. In contrast, for $T<T_o\approx 7.2$ K the respective
quantities are non-monotonic. In addition, for magnetic fields
$H>H_{\rm ''smp''}$ $S/T$ and $U_o/k_B$ present a monotonic
behavior as a function of temperature, while for $H<H_{\rm
''smp''}$ they are non-monotonic on $T$. As a consequence for
$H<H_{\rm ''smp''}$ $S/T$ and $U_o/k_B$ should present also
extrema as a function of the temperature. In all cases the
magnetic field was normal to the surface of the film ${\bf
H}\parallel{\bf c}$.}
\label{b7}%
\end{figure}%

The relaxation measurements performed in temperatures
$T>T_o\approx 7.2$ K showed different behavior. In Fig. \ref{b7}
we present comparative data for $S/T$ (upper panel) and $U_o/k_B$
(lower panel) at $T=2.5, 3.5, 4.3, 5$ K (below $T_o$) and at
$T=7.5, 8$ K (above $T_o$). In the upper panel we observe that the
difference $S/T\mid_{max}-S/T\mid_{min}$ of the two extrema (as
presented for example for the curve referring to $T=3.5$ K) is
higher at low temperatures. We may then conclude that as the
characteristic temperature $T_o\approx 7.2$ K is approached, the
extrema in $S/T$ and $U_o/k_B$ are suppressed and eventually
disappear at $T_o$. Furthermore, we observe that both $S/T$ and
$U_o/k_B$ exhibit opposite dependence on $T$ in the regimes below
and above the ''SMP'' points. For $H>H_{\rm ''smp''}(T)$ $S/T$
($U_o/k_B$) is an increasing (decreasing) function of temperature,
while for $H<H_{\rm ''smp''}(T)$ both $S/T$ and $U_o/k_B$ are
non-monotonic as a function of temperature. This means that for
$H<H_{\rm ''smp''}(T)$ both $S/T$ and $U_o/k_B$ exhibit extrema
not only as a function of field but also as a function of
temperature. In contrast, in the temperature regime above $T_o$
the behavior is completely different. As we see at $T=7.5$ and $8$
K both $S/T$ and $U_o/k_B$ present a conventional monotonic
behavior as they approach the upper-critical field $H_{\rm
c2}(T)$.

\section{conclusions}

In summary, we presented magnetic measurements as a function of
field, temperature and time in films of the isotropic Nb
superconductor prepared by annealing during the deposition. In
measurements as a function of field we observed a ''SMP'' feature,
while TMIs in the magnetic moment were observed either as a
function of field, temperature and time. The three lines $H_{\rm
fj}(T)$, $H_{\rm ''smp''}(T)$ and $H_{\rm fp}(T)$ connect at a
characteristic point $(T_o,H_o)\approx(7.2$ K,$80$ Oe). Below
$(T_1,H_1)\approx(6.2$ K,$1600$ Oe) the line $H_{\rm ''smp''}(T)$
changes slope. Pronounced TMIs are observed for $T_1\approx 6.2$
K$<T<T_o=7.2$ K, while below $T_1$ the magnetic curves are almost
regular. Interestingly, in this low-temperature regime $T<T_1$,
the first flux jump field preserves a constant value $H_{\rm
fj}(T<T_1)=40$ Oe. This is in strong disagreement to theoretical
proposals for thin film, or even bulk samples and remains to be
explained.

Relaxation measurements showed that below $T_o\approx 7.2$ K both
the activation energy $U_o$ and the normalized relaxation rate $S$
present a non-monotonic behavior either as a function of
temperature or the applied field. The observed extrema in $U_o$
and $S$ are located at the onset $H_{\rm on}(T)$ or the peak
points $H_{\rm ''smp''}(T)$ of the ''SMP''. This is a noticeable
difference comparing to high-$T_c$ superconductors where the
respective features are observed at a characteristic field
$H^*(T)$ placed well below the SMP point. This could give
additional evidence that the ''SMP'' observed in Nb films is of
different origin than the respective feature observed in
high-$T_c$ compounds. Although we are not able to reveal the
underlying mechanism that motivates the ''SMP'', our results
indicate that the ''SMP'' in Nb films is accompanied by an
anomalous relaxation of vortices. In contrast to high-$T_c$
compounds, in Nb films this anomalous relaxation could even be the
driving cause of the observed ''SMP''.

Although of different underlying physical mechanisms, the
similarities of the presented phase diagram of vortex matter of
our Nb disordered film with the phase diagrams observed in
high-$T_c$ compounds calls for further experimental and
theoretical investigation.

\begin{acknowledgments}
A. Speliotis should be gratefully acknowledged for useful
contribution in the preparation of samples.
\end{acknowledgments}

\pagebreak


\begin{references}


\bibitem{Esquinazi99} P. Esquinazi, A. Setzer, D. Fuchs, Y. Kopelevich,
E. Zeldov and C. Assmann, Phys. Rev. B 60 (1999) 12454.

\bibitem{Kopelevich98} Y. Kopelevich, and P. Esquinazi, J. Low Temp. Phys.
113 (1998) 1.

\bibitem{Stamopoulosnew} D. Stamopoulos, A. Speliotis, and D. Niarchos, Supercond. Sci. Technol. 17 (2004) 1261.

\bibitem{Kupfer94} H. Kupfer, S.N. Gordeev, W. Jahn, R. Kresse, R.
Meier-Hirmer, T. Wolf, A.A. Zhukov, K. Salama, and D. Lee, Phys.
Rev. B 50 (1994) 7016.

\bibitem{Zhukov95} A.A. Zhukov, H. Kupfer, G. Perkins, L.F. Cohen, A.D. Caplin, S.A. Klestov, H. Claus, V.I. Voronkova, T. Wolf and H. Wühl, Phys. Rev. B 51 (1995) 12704.

\bibitem{Janossy96} B. Janossy, L. Nguyen, and P. Wyder, Phys. Rev. B
53 (1996) 11845.

\bibitem{Abulafia96}Y. Abulafia, A. Shaulov, Y. Wolfus, R. Prozorov, L. Burlachkov, Y.
Yeshurun, D. Majer, E. Zeldov, H. Wühl, V.B. Geshkenbein, and V.
M. Vinokur, Phys. Rev. Lett. 77 (1996) 1596.

\bibitem{KhaykovichPRL} B. Khaykovich, E. Zeldov, D. Majer, T.W. Li, P.H. Kes, and M. Konczykowski Phys. Rev. Lett. 76 (1996) 2555.

\bibitem{Giller97} D. Giller {\it et al}., Phys. Rev. Lett. 79 (1997)
2542; D. Giller, A. Shaulov, Y. Yeshurun, and J. Giapintzakis,
Phys. Rev. B 60 (1999) 106.

\bibitem{Pissas99}M. Pissas, D. Stamopoulos, E. Moraitakis, G. Kallias, D. Niarchos,
and M. Charalambous, Phys. Rev. B 59 (1999) 12121.

\bibitem{Pissas00}M. Pissas, E. Moraitakis, G. Kallias, and A. Bondarenko, Phys. Rev. B 62 (2000) 1446.

\bibitem{Sun00} Y.P. Sun, Y.Y. Hsu, B.N. Lin, H.M. Luo, and H.C. Ku, Phys. Rev. B 61 (2000) 11301.

\bibitem{Miu00} L. Miu, E. Cimpoiasu, T. Stein, and C.C. Almasan, Physica C
334 (2000) 1.

\bibitem{Kokkaliaris00} S. Kokkaliaris, A.A. Zhukov, P.A.J. de Groot, R. Gagnon, L. Taillefer, T. Wolf, Phys. Rev. B
61 (2000) 3655.

\bibitem{Stamopoulos01}  D. Stamopoulos and M. Pissas, Supercond. Sci. Technol. 14 (2001) 844.

\bibitem{Stamopoulos02}  D. Stamopoulos and M. Pissas, Phys. Rev. B 65 (2002) 134524.

\bibitem{Stamopoulos03} D. Stamopoulos, M. Pissas, and A. Bondarenko, Phys. Rev. B 66 (2002) 214521.

\bibitem{Sun02}Y.P. Sun, W.H. Song, J.J. Du, and H.C. Ku, Phys.
Rev. B 66 (2002) 104520.

\bibitem{Chowdhury03}P. Chowdhury, Heon-Jung Kim, W.N. Kang, Dong-Jin
Zang, Sung-Ik Lee, and H. Kim, Phys. Rev. B 68 (2003) 134413; P.
Chowdhury, Heon-Jung Kim, In-Sun Jo, and Sung-Ik Lee, Physica C
384 (2003) 411.

\bibitem{Finnemore66} D.K. Finnemore, T.F. Stromberg, and C.A. Swenson,
Phys. Rev. 149 (1966) 231.

\bibitem{Mints96} R.G. Mints and E.H. Brandt, Phys. Rev. B 54 (1996) 12421.

\bibitem{Wipf6791} S.L. Wipf, Phys. Rev. 161 (1967) 404;{\it ibid} Cryogenics 31 (1991) 936.

\bibitem{Mints81} R.G. Mints and A.L. Rakhmanov, Rev. Mod. Phys. 53 (1981) 551.

\bibitem{Swartz68} P.S. Swartz and C.P. Bean, J. Appl. Phys. 39 (1968) 4991.

\bibitem{Anderson62}P. W. Anderson, Phys. Rev. Lett. {\bf 9},  (1962) 309;
P.W. Anderson and Y.B. Kim., Rev. Mod. Phys. 36 (1964) 39.

\bibitem{Gurevich94} A. Gurevich and E. H. Brandt, Phys. Rev.
Lett. 73 (1994) 178.


\end{references}
\end{document}